\documentclass[sigconf,nonacm, natbib=false]{acmart}

\AtBeginDocument{%
  }

\setcopyright{none}
\RequirePackage[
  datamodel=acmdatamodel,
  style=acmnumeric,
  ]{biblatex}

\usepackage{tabularx}
\usepackage{svg}
\usepackage{makecell}

\begin{document}
\title{Bias Mitigation for AI-Feedback Loops in Recommender Systems: A Systematic Literature Review and Taxonomy}

\author{Theodor Stoecker}
\affiliation{%
\institution{Technical University of Munich}
  \city{Munich}
  \country{Germany}
}
\email{theo.stoecker@tum.de}

\author{Samed Bayer}
\affiliation{%
\institution{Technical University of Munich \& Fraunhofer Gesellschaft}
  \city{Munich}
  \country{Germany}
}
\email{samed.bayer@tum.de}

\author{Ingo Weber}
\affiliation{%
  \institution{Technical University of Munich \& Fraunhofer Gesellschaft}
  \city{Munich}
  \country{Germany}
}
\email{ingo.weber@tum.de}

\renewcommand{\shortauthors}{Stoecker, Bayer, and Weber}

\begin{abstract}

Recommender systems continually retrain on user reactions to their own predictions, creating AI feedback loops that amplify biases and diminish fairness over time.
Despite this well-known risk, most bias mitigation techniques are tested only on static splits, so their long-term fairness across multiple retraining rounds remains unclear.
We therefore present a systematic literature review of bias mitigation methods that explicitly consider AI feedback loops and are validated in multi-round simulations or live A/B tests.
Screening 347 papers yields 24 primary studies published between 2019–2025.
Each study is coded on six dimensions: mitigation technique, biases addressed, dynamic testing set-up, evaluation focus, application domain, and ML task, organising them into a reusable taxonomy.
The taxonomy offers industry practitioners a quick checklist for selecting robust methods and gives researchers a clear roadmap to the field's most urgent gaps. Examples include the shortage of shared simulators, varying evaluation metrics, and the fact that most studies report either fairness or performance; only six use both.

\end{abstract}

\maketitle


\section{Introduction} \label{sec:introduction}

Recommender systems (RS) rely on dynamic machine learning (ML) models to personalise suggested content at scale. 
The algorithms continuously learn from the repeated interactions based on their own prior predictions~\cite{Pagan} or, more generally, outputs~\cite{AI-Engineering-Book-2025}.
This may lead to self-reinforcing AI-feedback loops, which, over successive cycles, can further amplify the prior bias in the system~\cite{huyen2025ai_engineering, klimashevskaia2024survey_popularity_bias}.
Such bias amplification harms the recommendation diversity, system fairness, long-term platform health, and user trust~\cite{mansoury2020feedbackloopbiasamplification}.

Most bias mitigation approaches for RS are evaluated on a single iteration of the training/validation/testing data splits, ignoring the feedback loop effect, therefore introducing evaluation bias~\cite{klimashevskaia2024survey_popularity_bias, Bauer2024survey_rs_evaluation}.
According to a related survey, 115 studies out of 127 are evaluated using offline testing without considering model updates~\cite{klimashevskaia2024survey_popularity_bias}.
Recent work in the literature, however, indicates a shift towards dynamic bias auditing. 
Simulations that replay many retraining rounds show that bias-mitigation approaches, which initially succeed, can fail to mitigate the bias in the long term~\cite{akpinar_long-term_2022}.
Frameworks such as FADE~\cite{yoo2024ensuringusersidefairnessdynamic} and FairAgent~\cite{guo2025enhancingnewitemfairnessdynamic} update models while enforcing fairness constraints.
Another study documents the AI-feedback loop amplification of source bias when RS learn from AI-generated content~\cite{zhou2025exploringescalationsourcebias}.

Although recent surveys classify individual bias categories and mitigation strategies, 
the field still lacks a systematic, empirical overview that links biases with dynamic mitigation strategies that remain effective under continual learning with feedback loops.
Our work addresses this gap with twofold contributions:
\begin{itemize}
    \item We conduct a systematic literature review on bias mitigation strategies within ML Model feedback loops, tested with retraining in simulation or live environments.
    \item Based on the literature, we propose a taxonomy that organises bias mitigation techniques in recommender systems by mitigation type, biases addressed, dynamic testing type, evaluation focus, application domain and ML model task.
\end{itemize}
In RS, feedback loops can lead to biased and unfair decisions that threaten the long-term health of platforms. Thus, our work has implications for both academic researchers and industry practitioners.

\section{Background and Related Work} \label{sec:related_work}

This section first outlines biases in RS and then summarises feedback-loop types, bias mitigation approaches, and existing surveys.

\subsection{Bias in Recommender Systems}
\label{sec:related_work_biases}

Machine learning models create predictions based on statistical patterns learned through observed data~\cite{bishop2006pattern,AI-Engineering-Book-2025}.
In RS, these models are updated based on new data collected in user interaction steps such as user feedback for suggested items~\cite{Pagan}. This leads to a loop of prediction, interaction, and retraining, where each step influences the others.

Suresh and Guttag categorise bias into seven types: \emph{historical}, \emph{representation}, \emph{measurement}, \emph{aggregation}, \emph{learning}, \emph{evaluation}, and \emph{deployment}~\cite{suresh_guttag_biases}. 
Moreover, Chen et al. present another influential taxonomy of bias types in RS with the feedback loop and seven classes: \emph{selection bias}, \emph{exposure bias}, \emph{conformity bias}, \emph{position bias}, \emph{inductive bias}, \emph{popularity bias}, and \emph{unfairness}
~\cite{chen_bias_and_debias_survey}.

\begin{figure*}[htbp]
  \centering    
  \includegraphics[width=1\textwidth]  {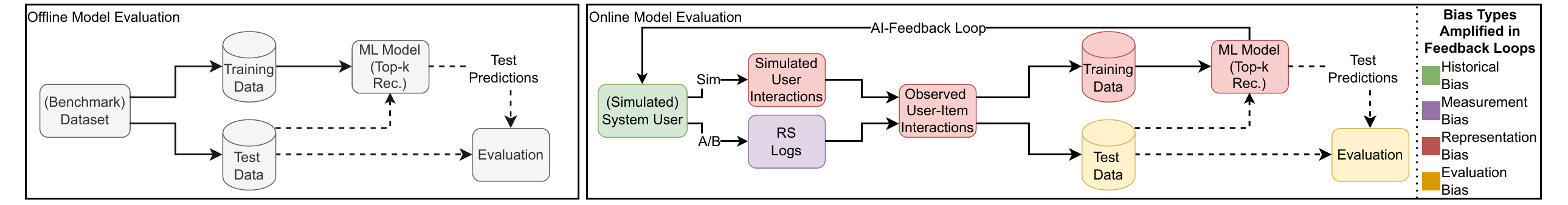}
  \caption{
  Offline evaluation detects bias in static data, whereas online (or simulated) evaluation reveals how feedback loops amplify four bias types---historical (green), measurement (purple), representation (red), and evaluation (yellow)---giving a more realistic view of model dynamics over time.
  }
  \Description{High level overview of the Offline and Online Model Evaluation pipeline.}
  \label{fig:ML_pipeline}
\end{figure*}

Following the feedback loop framework of Pagan et al. \cite{Pagan} (see Section \ref{sec:related_work_feedback_loops}), we decide to focus on four of Suresh and Guttag's bias classifications for our taxonomy: \emph{representation}, \emph{historical}, \emph{measurement}, and \emph{evaluation}.
These are the ones that are most relevant to feedback loops.
Figure \ref{fig:ML_pipeline} illustrates how evaluating a model in a (simulated) live setting exposes the recurring nature of these biases and how they can be amplified.

\subsection{Biases in AI-Feedback Loops} \label{sec:related_work_feedback_loops}
 Pagan et al. identified five types of feedback loops, characterised by their position within the ML system and the component affected: \emph{Sampling feedback loop}, \emph{Individual feedback loop}, \emph{ML Model feedback loop}, \emph{Feature feedback loop}, and \emph{Outcome Feature Loop}~\cite{Pagan}.

ML Model feedback loops arise when a system retrains (or evaluates) itself on the very instances it has already considered as only belonging to a specific class, such as the relevance for a user~\cite{Pagan}.
Figure~\ref{fig:ML_pipeline} shows how RS exemplify this concept: only recommended items receive user feedback.
Adding this feedback to the training data amplifies representation bias, whereas adding it to the test set reinforces evaluation bias. 
Additional loops can arise---for example, Individual feedback loops in which users modify their preferences in response to the system’s suggestions~\cite{Pagan}.

For the rest of this work, we differentiate the types of feedback loops based on their classification. 
We focus primarily on ML Model feedback loops in our literature review selection process, in order to ensure comparability of mitigation techniques. 
Unless explicitly naming a feedback loop type, feedback loops, including AI feedback loops, refer to ML Model feedback loops in later sections.

\subsection{Bias Mitigation Techniques and Classifications} \label{sec:related_work_mitigation_types}
Related works often use a well established framework for bias mitigation classification based on their stage in the ML-pipeline: \emph{Pre-}, \emph{In-}, or \emph{Post-Processing}. 
Pre-processing considers changes done before the model uses the data as input, in-processing describes an approach that changes some part of the prediction or learning process, and post-processing changes the model output~\cite{hort_pipeline_example,caton_pipeline_example,klimashevskaia2024survey_popularity_bias}.

Although the pipeline‐based taxonomy is well established, its concrete subclasses vary across the literature (cf. \cite{hort_pipeline_example,caton_pipeline_example}). 
Both Hort et al. and Caton and Haas concentrate on classifiers and therefore overlook tasks such as causal inference and reinforcement learning that are commonly used in RS.

We include Caton and Haas' classification despite their limitations for RS due to their impact in the related field of ML fairness and classifiers. 
They identify $16$ different classes within the three pipeline stages: \emph{Adversarial Learning}, \emph{Causal Methods}, \emph{Relabelling and Perturbation}, \emph{Resampling}, \emph{Reweighing}, \emph{Transformation}, and \emph{Variable Blinding} for \textbf{Pre-Processing}; \emph{Adversarial Learning}, \emph{Bandits}, \emph{Constraint Optimisation}, \emph{Regularisation}, and \emph{Reweighing} for \textbf{In-Processing}; and \emph{Calibration}, \emph{Constraint Optimisation}, \emph{Thresholding}, and \emph{Transformation} for \textbf{Post-Processing} \cite{caton_pipeline_example}.
Because many of those sub-classes either do not appear in or are not relevant to the feedback loop studies in RS, we introduce an additional set of sub-classes in the following sections.

\subsection{Existing Surveys on Biases in AI Feedback Loops}
We identify three prior surveys that also examine bias in RS, each from a different angle: a bias taxonomy for RS, a focused study on popularity bias, and a work on causal inference mitigation methods.

The closest related survey addresses seven different biases and the feedback loop effect in RS~\cite{chen_bias_and_debias_survey}. However, by developing distinct bias categories and adopting a single feedback loop concept, they offer a perspective that differs from the classification of feedback loops and biases outlined above.
Another study examines a single bias---popularity bias---but surveys a broader range of mitigation methods, many of which are evaluated only in offline settings~\cite{klimashevskaia2024survey_popularity_bias}. Consequently, its scope differs from ours in both the biases addressed and the evaluation approaches considered.
Lastly, Li et al. consider causal inference mitigation techniques for RS~\cite{li2025survey_causal_inference_recommenders}.
While we too included studies on causal inference techniques, we also include other approaches, as explained in the prior section.

Our research is different from existing works by focusing specifically on the current state of bias mitigation strategies for ML Model feedback loops, evaluated in a dynamic environment, including model updates.


\section{Method} \label{sec:methodology}
We queried two scientific databases: \textit{ACM Digital Library}, because it includes the most relevant conferences in the field, such as RecSys, WWW, or KDD; and \textit{IEEE Xplore} for their relevance in technical fields as a complementary source. Other databases were excluded due to resource constraints.
We added studies by two other methods: the inclusion of already identified ML Model feedback loop papers in the classification of Pagan et al.~\cite{Pagan}; and an additional search on ArXiv, to also include the most up-to-date research on the topic, despite their lack of a full peer review.

\begin{table*}[t]
  \centering
  \caption{Search details including the database, the last access time, the used search string, and the number of studies found.}
  \footnotesize
  \label{tab:search_strings}
  \begin{tabularx}{\textwidth}{@{}l c >{\raggedright\arraybackslash}X c@{}}
    \toprule
    \textbf{Database} & \textbf{Last accessed} & \textbf{Search string} & \textbf{Studies found} \\
    \midrule
    IEEE Xplore & 2025-06-13 &
      ("Abstract":"*feedback loop" OR
  "Abstract":"bias amplification" OR
  "Publication Title":"*feedback loop" OR
  "Publication Title":"bias amplification" OR
  "Author Keywords":"*feedback loop" OR
  "Author Keywords":"bias amplification")
AND
( "Full Text .AND. Metadata":mitigation OR
  "Full Text .AND. Metadata":guardrail*)
AND
("Abstract":bias OR
  "Abstract":biases OR
  "Publication Title":bias OR
  "Publication Title":biases OR
  "Author Keywords":bias OR
  "Author Keywords":biases)
    & 13 \\[2ex]
    ACM DL & 2025-05-23 & query: \{AllField:("Machine Learning" OR ML OR "Artificial intelligence" OR "AI" OR "Deep Learning") AND (Abstract:(bias*) OR Keyword:(bias*)) AND (Keyword:("*feedback loop*" OR "*bias amplification*" OR "*reinforc* feedback*" OR "echo chamber" OR "Recommender System") OR Abstract:("*feedback loop*" OR "*bias amplification*" OR "*reinforc* feedback*" OR "echo chamber" OR "Recommender System")) AND AllField:(Prevention OR mitigation OR "mitigation strategy" OR countermeasure OR control OR reduction)\}	"filter": 
     & 318 \\
     \midrule
    ArXiv & 2025-08-25 & order: -announced\_date\_first; size: 200; date\_range: from 2019-06-01 to 2025-08-31; classification: Computer Science (cs); include\_cross\_list: True; terms: AND all="machine learning" OR ML OR "artificial intelligence" OR AI OR "deep learning"; AND all=bias; AND all=feedback loop; AND all=mitigation&11\\
    \bottomrule
  \end{tabularx}
\end{table*}

Our inclusion criteria required 
i) the presence of an ML Model feedback loop (see Section~\ref{sec:related_work_feedback_loops}), and 
ii) an applied approach of bias mitigation, tested in 
iii) a dynamic environment with updates of the model, such as in simulations or live-testing with more than one iteration of a loop, such as depicted in Figure \ref{fig:ML_pipeline}.
Only research papers from conferences, workshops, and journals were considered; extended abstracts, and posters were excluded.

\begin{figure}[htbp]
  \centering    
  \includegraphics[
  width=1.0\columnwidth
  ]{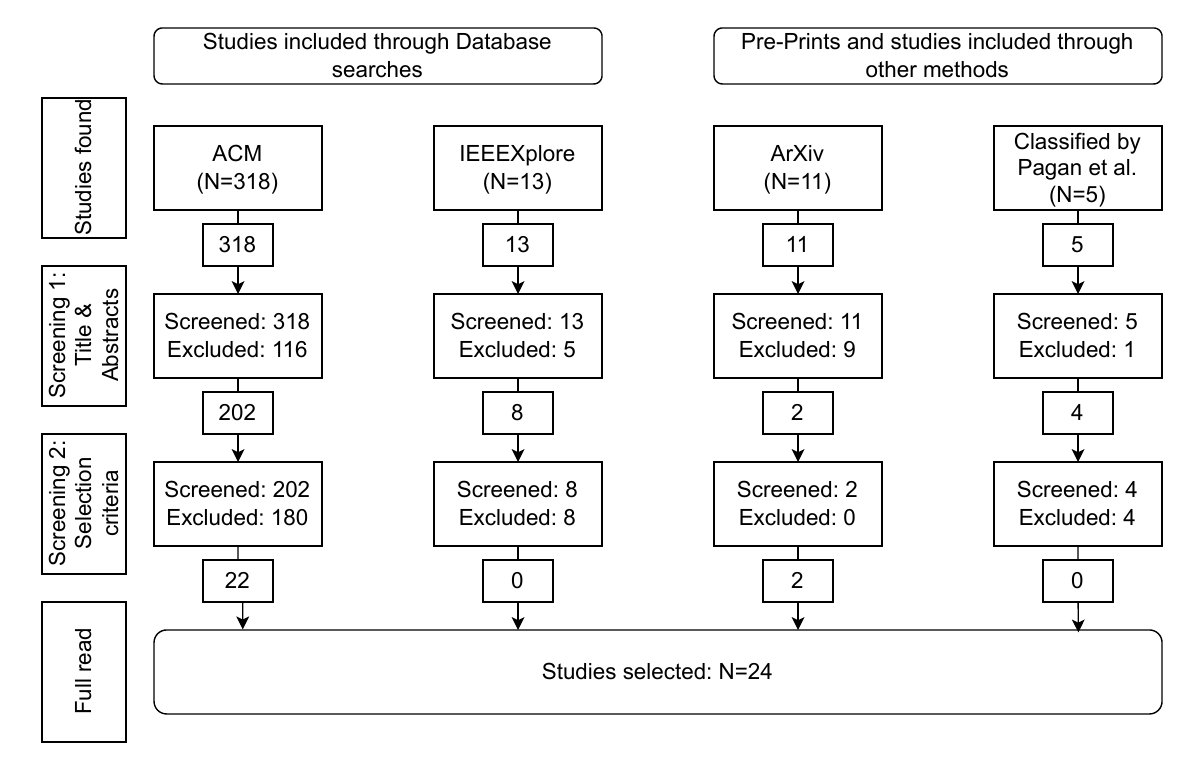}
  \caption{
  Flowchart of study selection. From $347$ records identified (ACM $318$; IEEE Xplore $13$; ArXiv $11$; Pagan et al. $5$), titles and abstracts were first screened for AI-feedback loops and bias mitigation; full texts were then assessed for applied mitigation strategies tested via simulation or A/B testing, yielding $24$ studies.
  }
  \Description{Number of selected papers in each stage.}
  \label{fig:prisma_funnel}
\end{figure}

The search strings shown in Table~\ref{tab:search_strings} were chosen to find relevant studies by finding ML related research on bias mitigation in feedback loops. 
As fields use different words to describe a similar phenomenon, we also included "echo chamber" to also consider related studies. 
This same search string, however, did not yield any papers for IEEE Xplore, so we queried a search without the explicit mention of ML or related words.
For the ArXiv search, we conducted a simple search for feedback loops in ML in connection with bias mitigation published in the last six years.
The most common reasons for exclusion were missing AI-feedback loops in screening stage 1 and a lack of an online evaluation in screening stage 2.

Our procedure to select relevant papers as outlined in Figure~\ref{fig:prisma_funnel} was twofold. In a first screening, we searched for relevancy by examining title and abstract for a relation to AI-feedback loops, biases, and mitigation strategies. This led to $150$ papers.  
To augment our selected texts and to account for personal bias, we utilised a large language model (LLM: \emph{gpt-4o-mini}\footnote{
    See prompt used in \href{https://docs.google.com/spreadsheets/d/11pklSwOYsv6PbalBjLUBoGdestKsSl2v_dpS50i3sgw/edit?usp=sharing}{Online Appendix}
}) for texts from the ACM database, which identified $158$ papers in this screening step (93 overlapping with our initial set). This yielded 216 non-duplicate papers for full-text assessment.
The final selection decision remained human in the next step.
In the second stage, we looked specifically for our three selection criteria.
After full-text review, 24 papers met all criteria. A second evaluator independently checked this selection for consistency.

We constructed the taxonomy using Nickerson et al.'s framework through multiple conceptual-to-empirical iterations~\cite{nickerson2013method}.
First, we began with established categories from prior surveys or related work (Section \ref{sec:related_work}). 
We then iteratively mapped every candidate study onto the current taxonomy.
When a study did not fit clearly to the current categories in the literature,
we tailored category definitions or introduced a new sub-class (e.g., Add-On for architectural extensions).
We ended the iterative process once all studies fit into the taxonomy, satisfying the objective and subjective stopping criteria outlined by Nickerson et al.~\cite{nickerson2013method}.
Finally, we coded the full set of 24 studies with the final taxonomy in Table \ref{tab:papers_core_dimensions}.


\section{Results} \label{sec:results}
We start by presenting metadata about the studies, and continue with the criteria outlined in Section~\ref{sec:methodology}. The list of papers mapped to our criteria can be found in our table for reproducibility \footnote{
    See full list of papers in the
    \href{https://docs.google.com/spreadsheets/d/11pklSwOYsv6PbalBjLUBoGdestKsSl2v_dpS50i3sgw/edit?usp=sharing}{Online Appendix}
    }.

\subsection{Study Metadata and Attributes}

\begin{figure}[htbp]
  \centering
  \includegraphics
  [width=1.0\columnwidth]
  {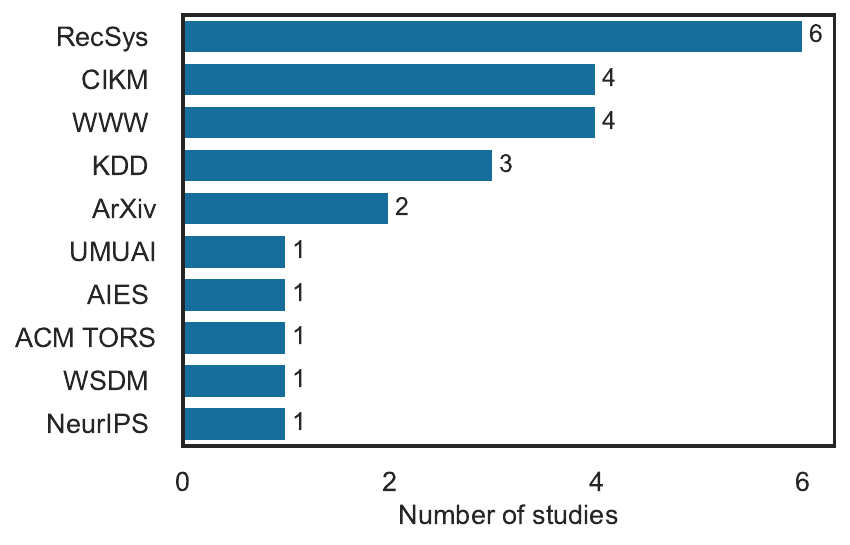}
  \caption{Number of included studies per publication venue ($N=24$). RecSys leads with six studies (25~\%), followed by CIKM and WWW with four each (17~\%).}
  \Description{Included studies per publication venue.}
  \label{fig:venues_bar}
\end{figure}

Figure \ref{fig:venues_bar} lists the publication venues, most of which are implementation oriented outlets. Across these venues, $22$ of the $24$ papers in our sample of ML Model feedback loop studies address RS.
ACM's Conference on Recommender Systems (RecSys) is the most prominent one.  
Venues for fairness research, such as the ACM Conference on AI, Ethics, and Society (AIES), play only a smaller role.
Just two of the included studies were published in a journal, which indicates the fast pace and conference-focused nature of the field.

\begin{figure*}[htbp]
  \centering
  \includegraphics [width=1.0\textwidth]
  {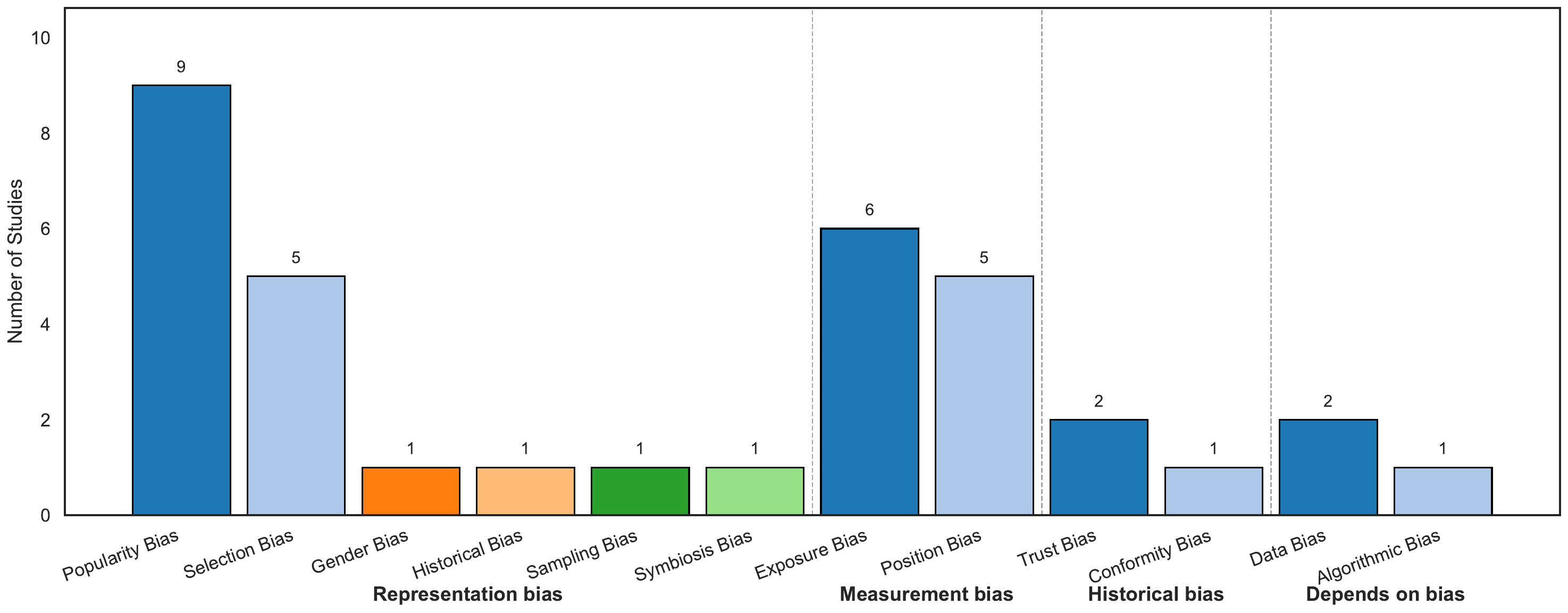}
  \caption{Biases mentioned and their mapping to Suresh and Guttag's framework. Data and algorithmic bias depend on the specific bias as these are too general terms to clearly classify. Please note that the representation category in a feedback loop can also lead to evaluation bias instead.}
  \Description{Bias types.}
  \label{fig:biases_mapped}
\end{figure*}

$17$ studies have at least one author affiliated with industry (71~\%), with $13$ of them having the majority of authors affiliated with industry (54~\%).
The extent of industry research, primarily from large platforms utilising RS, highlights the high demand in practice.
Co-authored studies of industry and academia also indicate the close cooperation between them.

The publication years ranged from 2019 to 2025, with 2020 and 2022 having the most studies, five and six, respectively.

\subsection{Domain of System Applications} \label{sec:results_domains}
While most of the papers considered general-purpose RS ($9$), the most popular domains are music ($2$), and movie/video recommendation ($4$). 
Domains of RS represented by only one paper fall into the “Other” category: E-Commerce, advertisements, social networks, app store, news, search recommendation, and A/B testing for recommendation systems.
The two studies unrelated to RS focused on reinforcement learning, and on loan applications, healthcare, and policing.
The variety of fields where RS are applied shows the applicability of mitigation approaches across different fields.

The different input data, ranging from text in news recommendation to music and video data, also highlights the need to find easily usable features for the RS. 
Examples of this are watch time, to extract the user interest of a video based on an easily measurable metric~\cite{lin_tree_2023}.

\subsection{ML-Model tasks} \label{sec:results_type_and_tasks}
The tasks for RS ranged from predicting specific metrics, candidate generation, to ranking the outputs.
As mentioned in the prior section, two systems are completely independent of RS, their tasks focused on classification and optimisation of resource allocation between groups.

Top-$k$ recommendation is the dominant task, addressed in $20$ studies. Of these, seven concentrate on the ranking phase, and two treat the special case of $k = 1$---where user feedback differs, as preferences over a list of items are absent when only a single item is presented.

Another type of tasks has the aim of predicting features used in the RS (e.g., Click through rate (CTR)~\cite{guo_pal_2019}, or watch time prediction~\cite{lin_tree_2023}). 
CTR and watch time are logged but unobserved at inference. By predicting these values from history, the model can treat them as features and improve the following top-$k$ recommendation.

\subsection{Biases addressed}

Studies were motivated by different, sometimes multiple, biases. 
For each paper, we recorded every bias that was explicitly addressed. For papers that did not name a specific bias, we inferred the nearest matching bias from the problem statement when possible.

Apart from the five biases identified in the survey on biases in RS by Chen et al. \cite{chen_bias_and_debias_survey}, an additional seven were named explicitly in the studies. The mentioned biases and their mapping to the framework of Suresh and Guttag~\cite{suresh_guttag_biases} are visualised in Figure~\ref{fig:biases_mapped}. "Depends on bias" is used in this mapping whenever the mentioned bias is too broad to be classified. Furthermore, note that in the one case of historical bias, the description of the bias used in the study is different from its description in the classification framework~\cite{suresh_guttag_biases}.

\begin{figure*}[htbp]
   \centering    \includegraphics[
  width=1.0\textwidth
  ]{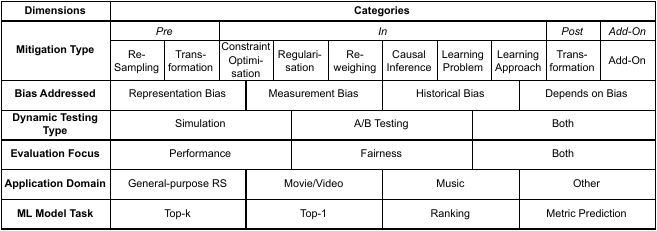}
  \caption{Taxonomy for recommender systems bias mitigation evaluated in dynamic environments based on six dimensions.
  }
  \Description{Proposed Taxonomy.}
  \label{fig:taxonomy}
\end{figure*}

\begin{figure}[htbp]
  \centering
  \includegraphics [width=1.0\columnwidth]
  {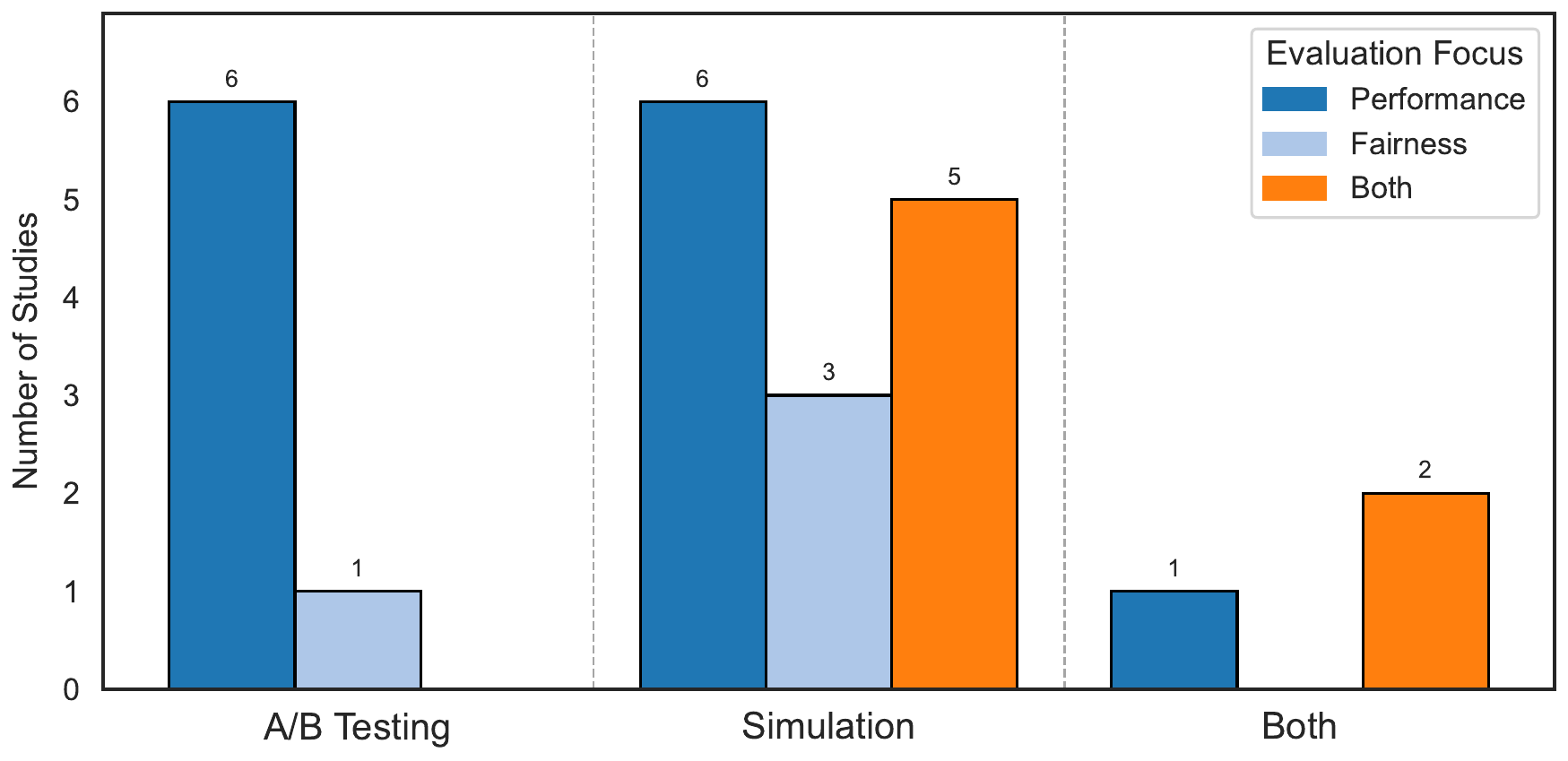}
  \caption{Metrics used to evaluate a mitigation type in simulated or A/B tested feedback loop environments. A/B tests mostly use performance type evaluations.}
  \Description{Evaluation metrics and dynamic environment types.}
  \label{fig:evaluation_focus_by_simulation_type}
\end{figure}

\subsection{Mitigation Types} \label{sec:results_mitigation_types}
Mitigation approaches relied on changes to the in-processing stage of a model in $17$ out of $24$ ($71$~\%) cases, the pre-, and post-processing stage were changed in two and three studies respectively.
Two studies changed the model or system architecture to mitigate biases, which is not covered in the established pre-/in-/post-processing.
We categorised them as "Add-On" as a separate mitigation type in this study.

Of the presented subclasses identified by Caton and Haas \cite{caton_pipeline_example}, we identified six in the context of RS:
 \emph{Resampling}, and \emph{Transformation} in \textbf{Pre-Processing}; \emph{Constraint Optimisation}, \emph{Regularisation}, and \emph{Reweighing} for \textbf{In-Processing}; and \emph{Transformation} for \textbf{Post-Processing}.
 
 To cover the remaining studies, we extend the taxonomy with four RS-specific classes: \emph{Causal Inference-based} from the survey on popularity bias in RS \cite{klimashevskaia2024survey_popularity_bias}, \emph{learning approach} to capture a change to a zero-shot method, and \emph{learning problem} for reinforcement-learning or reward or loss function changes \textbf{In-Processing}; and \emph{Add-On}, which covers the addition of components rather than changing the established pipeline steps, such as adding a Monte Carlo simulation.

\subsection{Evaluation and Metrics}
We visualised evaluation types and used metrics in Figure~\ref{fig:evaluation_focus_by_simulation_type}.
In the literature retraining is done by either using a simulation or live A/B tests.
In simulations, a user-system-interaction is simulated, whereas in an A/B test, two groups of live users get recommendations from different versions of the RS (one with bias mitigation, the other one without, as a control group).
54~\% of studies focused only on performance, 17~\% on measuring the effect of their mitigation technique using a fairness metric, and the other 29~\% investigated both performance and fairness. A/B Tests and industry works were most likely to be evaluated with performance metrics.

The most popular performance metrics were Click Through Rate (CTR)~\cite{gupta_causer_2021,klimashevskaia_evaluating_2023,guo_pal_2019,pessimistic_jeunen_2021,jeunen_pessimistic_2023,wu_zero-shot_2020}, and normalised Distributed Cumulative Gain (nDCG)~\cite{wu_zero-shot_2020,wang_off-policy_2022,ma_off-policy_2020} or variations of those. Other metrics included some commonly used in machine learning, such as mean absolute/squared error (MAE/MSE)~\cite{jeunen_pessimistic_2023,lin_tree_2023}, Area under the receiver operator curve (ROC-AUC)~\cite{guo_deep_2020}, or cumulative gain (for reinforcement learning)~\cite{tang_bandit_nodate}.
In one case, the performance metrics were adapted to specifically investigate the bias. Han et al. measured the overall performance, and newer items in particular, to see the effect on the cold-start problem~\cite{han_addressing_2022}.

Evaluations based on fairness measurements vary by use case. 
GINI Coefficient was the most popular choice~\cite{ferraro_its_2024,ferraro_exploring_2020,chang_cluster_2024}. 
Other than that, specific metrics are used, for instance: difference in granted loans between gender~\cite{she_fairsense_2025}, or the position of first female artist for music recommendation~\cite{ferraro_its_2024}.


\section{Taxonomy and Discussion} \label{sec:discussion}

\newcolumntype{P}{>{\hsize=1.1\hsize\arraybackslash}X}
\newcolumntype{E}{>{\hsize=0.8\hsize\arraybackslash}X}
\newcolumntype{Q}{>{\hsize=0.8\hsize\arraybackslash}X}
\newcolumntype{D}{>{\hsize=0.5\hsize\centering\arraybackslash}X}

\begin{table*}[ht]
  \centering
  \footnotesize
  \setlength{\tabcolsep}{3pt}
  \renewcommand{\arraystretch}{0.8}
    \caption{Taxonomy mapping for each of the 24 papers across our seven core dimensions.  
    Biases in brackets are mapped in cases without explicitly mentioned biases. The last two studies are the ones unrelated to recommender systems. Representation, measurement, and historical bias are abbreviated as Rep., Meas., and Hist., respectively.
  }
  \label{tab:papers_core_dimensions}
  \begin{tabularx}{\textwidth}
  {l X X D Q P Q}
    \toprule
    \textbf{Paper} &
    \textbf{Mitigation Type} &
    \textbf{Bias Addressed} &
\textbf{\makecell{Dynamic\\Testing\\Type}} &
    \textbf{Evaluation Focus} &
    \textbf{Application Domain} &
    \textbf{ML Model Task} \\
    \midrule
\cite{han_addressing_2022}  &  Transformation (Pre) & Rep., Meas., and Hist. Bias & Both & Both & Other (E-Commerce) & Ranking\\
    \cite{gupta_causer_2021} & Causal Inference-based & Rep. Bias & Sim & Performance & General-purpose RS & Top-k\\
    \cite{chang_cluster_2024} &  Regularisation & Rep. Bias & A/B & Fairness & General-purpose RS & Top-k\\
    
    \cite{damak_debiasing_2022} & Reweighing & Meas. and Rep. Bias & Sim & Both & General-purpose RS & Top-k\\
    
    \cite{guo_deep_2020} & Learning Problem  & Depends & Both & Performance & Other (Advertisement) & Top-k\\
    \cite{capan_dirichletluce_2022} & Learning Problem  & Meas. and Rep. Bias & Sim & Both & General-purpose RS & Top-k\\
    \cite{klimashevskaia_evaluating_2023} & Transformation (Post) & Rep. Bias & A/B & Performance & Movie/Video & Top-k\\
    \cite{ferraro_exploring_2020}& Transformation (Post) & Rep. Bias & Sim & Both & Music & Top-k\\
    \cite{ferraro_its_2024} & Transformation (Post) & Rep., and Meas. Bias & Sim & Fairness & Music & Ranking\\
    
    \cite{akpinar_long-term_2022} & Constraint Optimisation & Rep., Meas., and Hist. Bias & Sim & Fairness & Other (Social Network) & Top-k\\
    
    \cite{su_multi-task_2024} & Learning Problem  & Meas. Bias & A/B & Performance & General-purpose RS & Ranking\\
    
    \cite{ma_off-policy_2020} &  Reweighing & (Meas. Bias) & Sim & Performance & General-purpose RS & Ranking\\
    
    \cite{wang_off-policy_2022} & Reweighing & Meas., Hist., and Rep. Bias & Sim & Performance & General-purpose RS & Top-k\\
    
    \cite{guo_pal_2019} & Add-On & Meas. Bias & A/B & Performance & Other (App Store) & Metric prediction\\
    \cite{jeunen_pessimistic_2023} & Learning Problem  & Rep. Bias & Sim & Performance & General-purpose RS & Top-1\\
    \cite{pessimistic_jeunen_2021} & Learning Problem  & Rep. Bias & Sim & Performance & General-purpose RS & Top-1\\
    \cite{brennan_reducing_2025}  & Resampling & Rep. Bias & Sim & Fairness & Other (A/B Testing for RS) & Ranking\\
    
    \cite{chen_top-k_2019} & Reweighing & Depends & Both & Both & Movie/Video & Top-k\\
    
    \cite{lin_tree_2023} & Causal Inference-based & Rep. Bias & A/B & Performance & Movie/Video & Metric prediction\\
    
    \cite{ren_unbiased_2022} & Reweighing & Meas., and Hist. Bias & A/B & Performance & Other (News) & Ranking\\
    
    \cite{wu_zero-shot_2020} & Learning Approach & (Rep. Bias) & A/B & Performance & Other (Search) & Top-k\\

    \cite{khenissi2020modelingcounteractingexposurebias} & Regularisation & Meas. Bias & Sim & Both & Movie/Video & Ranking\\
    
    \midrule
    \cite{tang_bandit_nodate} & Learning Problem & Rep. Bias & Sim & Performance & Non Recommender System & Other (Optimisation) \\
    
    \cite{she_fairsense_2025} &  Add-On & Depends & Sim & Both & Non Recommender System & Other (Classification)\\
    \bottomrule
  \end{tabularx}

\end{table*}

\subsection{Taxonomy}
As described in Section~\ref{sec:methodology}, we followed a well-established method~\cite{nickerson2013method} to derive a taxonomy -- the result is visualised in Figure \ref{fig:taxonomy}.
Our six-dimensional taxonomy captures bias-mitigation approaches for RS that incorporate an ML Model feedback loop and are evaluated in dynamic environments.
Table \ref{tab:papers_core_dimensions} classifies the $24$ studies from our literature review using the proposed taxonomy.

First, the mitigation type, rooted in the established pipeline-stage classification and its sub-categories, indicates both where in the workflow the bias-reducing intervention takes place and what it specifically modifies.
Next comes the bias type---the specific bias the authors highlight as their primary motivation.
While related taxonomies covered additional classes of mitigation types \cite{caton_pipeline_example,hort_pipeline_example}, they were unable to cover all identified types. We added four classes to classify those papers, as described in Section \ref{sec:results_mitigation_types}.
The lack of the other classes in their studies might indicate potential gaps in literature, but could also be due to the number of selected studies fitting our selection criteria, as well as the difficulty to implement specific types, such as adversarial learning approaches, for RS.
A third dimension tracks reiteration or dynamic testing, revealing how the model tries to show the effectiveness over a number of feedback loop iterations. 
Fourth, the evaluation focus tells us which metric is used to judge whether the performance of the changed model or system actually improves.
Finally, we distinguish between the domain, the real-world application in which the work operates, and the ML model task, that is, what the model ultimately predicts in RS.

This taxonomy allows the classification of new bias mitigation approaches while staying compatible with established bias and feedback loop frameworks outlined in earlier sections.

\subsection{Discussion}

When considering biases in RS, it is important to note that feedback loops amplifying the popularity of items can also be beneficial---a high-quality item might be suggested to a larger number of users. The challenge here lies in identifying the origin of popularity: quality or system-introduced~\cite{Zhang_Causal_benefit_bias}.

While ML Model feedback loops primarily consider representation and evaluation bias, some of the mentioned biases are classified as measurement or historical bias as well. This is because of three main reasons: 
Firstly, we confirm the finding of other studies that the usage of bias terminology is unclear within the literature~\cite{klimashevskaia2024survey_popularity_bias,Pagan,chen_bias_and_debias_survey}. 
In addition to this, the existence of different feedback loop types within one system, and the mention of multiple biases within a single study often obscures which specific biases are actually being examined.
Lastly, for the bias mapping we rely on the used bias terminology, not on a single study basis.


\section{Conclusion, Limitation and Future Work} \label{sec:conclusion}

In summary, bias mitigation in feedback looped recommender systems is a growing field, dominated by in-processing mitigation techniques. We conducted a systematic literature review yielding $24$ recent studies and introduced a taxonomy that helps categorise and compare bias mitigation techniques within feedback loops across six dimensions.

Our work is limited by the choice of mitigation approaches considered. 
Specifically, the databases used and the need for dynamic testing in the evaluation of the mitigation approach.
Another limitation concerns the databases we used; in future work,
further databases could be included to get a more holistic view of the topic.

Assuming the relations of evaluation methods applied found by Klimashevskaia et al.\ can roughly be applied to other biases, our exclusion criteria demanding live or simulation based evaluation of their approach excludes around 90~\% of studies~\cite{klimashevskaia2024survey_popularity_bias}. 
This indicates either a large gap between evaluations of popularity bias and the broader RS and feedback loop scope, or limits our work.

To also consider future studies, our taxonomy might need to be extended to accurately capture additional mitigation types.
Specifically, the general use of the learning problem category for reinforcement learning can be further refined.

The lack of widely applied simulation frameworks, such as a flexible benchmarking environment, indicates a potential direction for future research. 
This could increase comparability and further foster the already strong connection between academia and industry. 
Therefore, further assisting research should be directed at designing \textit{responsible feedback loop systems}---ensuring both fairness to users, as well as the long-term health of recommender systems.
Such efforts would guide academic exploration and facilitate practical adoption in real-world platforms.

\printbibliography

\end{document}